\documentclass[epj]{svjour}
\usepackage{graphics}
\begin{document}

\headnote{Short Note}

\title{$^{\rm 7}$Li quasi-free scattering off the $\alpha$-cluster
 in $^{\rm 9}$Be nucleus}

\author{ N. Soi\'{c}\inst{1} \and D. Cali\inst{2} \and
 S. Cherubini\inst{2}\thanks{\emph{Present address:} Institut de Physique Nucl\'{e}aire, 
Universit\'{e} Catholique de Louvain, Louvain-la-Neuve, Belgium }
 \and E. Costanzo\inst{2} \and M. Lattuada\inst{2} \and M. Milin\inst{1}
 \and \mbox{D\raisebox{0.30ex}
{\hspace*{-0.75em}-}\hspace*{ 0.42em}}. Miljani\'{c}\inst{1}
 \and S. Romano\inst{2} \and C. Spitaleri\inst{2} \and M. Zadro\inst{1} }

\institute{ Ru\mbox{d\raisebox{0.75ex}
{\hspace*{-0.32em}-}\hspace*{-0.02em}}er
Bo\v{s}kovi\'{c} Institute, Zagreb, Croatia
 \and INFN-Laboratori Nazionali del Sud and
 Universit\`a di Catania, Catania, Italy }

\date{Received: 1. Jun 1998. / Revised version: 1. Oct 1998.}
%
\abstract{
{ Spectra of coincident charged particles from the reactions
 induced by a 52 MeV $^{7}$Li beam incident on a beryllium target were
 measured. Strong contributions of the $^{7}$Li quasi-free scattering off
 the $\alpha$-cluster in $^{9}$Be nucleus were observed. This
 observation supports the conclusions from the study of complete fusion
 of weakly bound light nuclei at low energies that the "fragility" of
 the nuclei makes their fusion less probable.
}
 }
%
\maketitle

 $^{7}$Li{} and $^{9}$Be{} nuclei are very deformed, loosely bound systems. 
 In the cluster model their ground states are portrayed as having the 
 $\alpha{}$-t and $\alpha{}$-$^{5}$He cluster structure, respectively.
 It has been recently found \cite{1}, that the cross section for fusion
 of light, weakly bound nuclei at energies ranging from E$_{\rm C}$ to
 5E$_{\rm C}$ (E$_{\rm C}$ - Coulomb barrier) is significantly lower than
 the total reaction cross section and also smaller than the fusion cross
 section expected from the available systematics. This feature was
 interpreted to be due to the strong influence of the break-up processes.
 Among the break-up processes, especially at higher energies, an important
 role is often played by the quasi-free scattering, a process in which one
 of the nuclei in collision quasielastically scatters off a cluster and
 ejects it from the other nucleus. Many different quasi-free processes
 have been observed and studied with both $^{7}$Li{} and $^{9}$Be{}
 nuclei - from e.g. (p,p$\alpha${}) and ($\alpha${},2$\alpha${})
 reactions on them as target nuclei \cite{2,3,4} to their direct
 fragmentation as projectiles in interaction with complex nuclei
 e.g. \cite{5}. However, as far as we know, there is no information about
 these processes in reactions between these two "fragile" nuclei.
 Recently we have studied different many-body exit channels of the
 $^{9}$Be{}-$^{7}$Li{} reaction at $^{7}$Li{} energy of 52 MeV \cite{6}.
 This short note is concerned primarily with the observation of strong
 contributions of the $^{7}$Li{}
 quasi-free scattering on the $\alpha${}-particle cluster in 
 $^{9}$Be{} nucleus at this relatively low energy.

 In the experiment a 52 MeV $^{7}$Li$^{+++}$ beam (I = 60 - 100 nA) from
 the SMP Tandem Van de Graaff accelerator (Laboratori Nazionali del Sud)
 was used to bombard a self-supported beryllium target
 (400 $\mu$g cm $^{-2}$). Outgoing charged particles were detected and
 identified in several particle telescopes consisting either of silicon
 surface barrier ($\Delta${\it E} and {\it E}) detectors (SDT),
 or of an ionization chamber($\Delta${\it E}) and position sensitive silicon
 detector (E),(ICPSDT). The angular range covered by the ICPSDTs was
 8$^{\rm o}$,
 while the angular opening of the SDTs was 1$^{\rm o}$. Coincidence events
 between any two telescopes of different type were recorded.

\begin{figure}
\resizebox{0.5\textwidth}{!}{
  \includegraphics{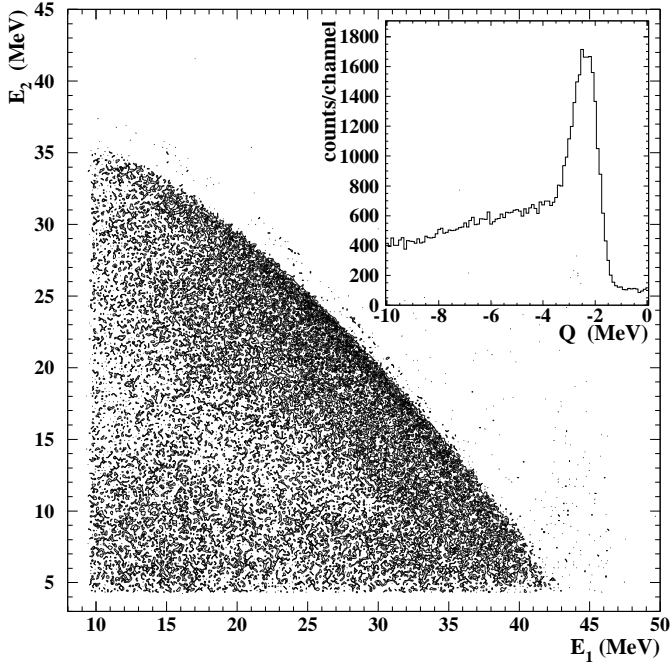}
}
\caption{Two-dimensional E$_{1}$-E$_{2}$ plot for the
 $^{9}$Be{}($^{7}$Li{}, $^{7}$Li{}$\alpha${})$\alpha${}n events measured at 
 E$_{0}$ = 52 MeV and for $\bar{\Theta}_{^{7}Li}$ = 26.8$^{\rm o}$ 
 $\Theta_{\alpha}$ = 40.3$^{\rm o}$. The Q-value spectrum constructed from
 these events is shown in the insert.  }
\label{fig:1}       
\end{figure}

 Fig. 1. shows the $^{7}$Li{}-$\alpha${} coincidence data for the ICPSDT
 positioned at $\Theta _{^{7}Li}$ = 26.8$^{\rm o}$ and for the SDT at
 $\Theta _{\alpha}$ = 40.3$^{\rm o}$
 ($\Delta \Phi _{ij}$ = 180$^{\rm o}$ for all spectra presented here). 
 The axes correspond to the total
 energies of particles from the reaction. An intense concentration of
 events can be observed on the boundary corresponding to the undetected
 $\alpha${}-particle and neutron left in the $^{5}$He ground state.
 Three weaker partially overlapping bands also show through the
 four-body (2 $\alpha${} + $^{7}$Li{} + n) continuum. They correspond to
 different processes in which one of the products
 ($^{5}$He$_{\rm 0}$, $^{8}$Be$_{\rm 0}$ and $^{8}$Li$_{\rm 2}$)
 decays with small energy 
 and either $\alpha${} or $^{7}$Li{} from the decay are detected.
 In the insert one can see the same data presented as 
 a Q-value spectrum, which evidently shows that the largest contribution
 to the $^{9}$Be{}($^{7}$Li{}, $^{7}$Li{}$\alpha${})$\alpha${}n reaction
 for this angle pair corresponds to the undetected particles being in the
 $^{5}$He ground state (Q $\approx$ - 2.5 MeV).

\begin{figure}
\resizebox{0.5\textwidth}{!}{
  \includegraphics{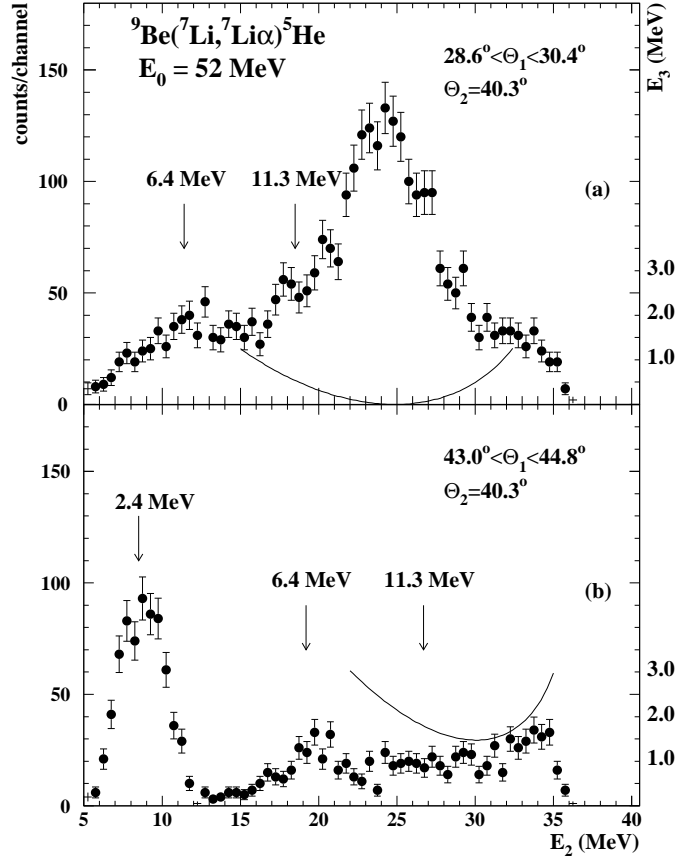}
}
\caption{Projections onto the $\alpha${}-particle energy axis of
 the $^{7}$Li{}-$\alpha${} coincident spectra for
 $\Theta_{\alpha}$ = 40.3$^{\rm o}$ and $\Theta_{^{7}Li}$ =
 29.5$^{\rm o}$ (a) and $\Theta_{^{7}Li}$ = 43.9 (b). The spectra are
 constructed from the events having Q = ( -2.45 $\pm$ 0.55) MeV.
 The curves represent energies of undetected $^{5}$He nucleus. }
\label{fig:2}   
\end{figure}

  Fig. 2. presents the $^{7}$Li{}-$\alpha${} coincident spectra from
 the $^{9}$Be{}($^{7}$Li{}, $^{7}$Li{}$\alpha${})$\alpha${}n reaction,
 measured for $\Theta _{\alpha}$ = 40.3$^{\rm o}$ and
 $\Theta _{^{7}\rm Li}$ = 29.5$^{\rm o}$ and 43.9$^{\rm o}$, and shown
 as projections on the $\alpha${}-particle energy axis. These
 spectra are formed from all those events satisfying the following 
 condition: Q = (-2.45 $\pm$ 0.55) MeV. The curves are the energies 
 of undetected $^{5}$He (right scale)
 and the arrows indicate positions where the
 contributions from some particle decaying states of 
 $^{9}$Be{} are expected. The prominent peak observed for
 $\Theta _{^{7}\rm Li}$ = 29.5$^{\rm o}$ and very low energies of 
 $^{5}$He{} ($<$ 800 keV) is due to the $^{7}$Li{} quasi-free 
 scattering off the $\alpha${}-cluster in $^{9}$Be{}. Similar
 broad peaks are observed for all angle pairs satisfying kinematical
 conditions for this process (E$_{3}^{\rm L}$ $\approx$ 0). In these cases
 they represent a significant fraction of the yield and the only other 
 contribution in the $^{7}$Li{}-$\alpha${} spectra with comparable 
 intensity is due to a sequential process as seen in the spectrum for 
 $\Theta _{^{7}\rm Li}$ = 43.9$^{\rm o}$. The process is $^{7}$Li{}
 inelastic scattering on $^{9}$Be{} leading to the 5/2$^{-}$ state,
 the second member of the ground state rotational band of $^{9}$Be{},
 which then decays by neutron emission into the tail of the broad 
 first excited state of $^{8}$Be{}. Weak contributions from the
 sequential process through the $\alpha{}$-$^{5}$He{} decaying state
 in $^{9}$Be{} at energies of 6.4 and 11.3 MeV can also be seen.

 The quasi-free peaks reflect in a complex way the momentum distributions
 of the knocked-out clusters in nuclei before the collision. In the
 factorized distorted-wave impulse approximation (DWIA), the theory
 most often used in the analysis of the quasi-free reactions, the
 cross section is essentially a product of two terms (additional terms
 being spectroscopic and kinematic factors): the so called distorted momentum 
 distribution for the knocked cluster - spectator relative motion and the
 half-off-shell cross section for the projectile - cluster interaction.
 The first term in the plane wave limit corresponds to the momentum 
 distribution of the cluster in the target nucleus. The half-off-shell
 cross section is most often replaced by experimental free projectile -
 cluster scattering cross section. In the present case although there
 exist some data on $^{7}$Li{}-$\alpha{}$ scattering \cite{7}, they
 could not be used because they cover an angular range different from
 the one in the experiment. One can expect that the distortions are
 severe at these low energies and with complex projectile like $^{7}$Li{}.
 Because of these and other ambiguities it did not seem worthwhile to
 perform elaborate DWIA calculations. However, for an illustration the
 $^{9}$Be{}($^{7}$Li{}, $^{7}$Li{}$\alpha${}) reaction cross section data
 for several angle pairs close to quasi-free scattering conditions,
 divided by the kinematical factor, are displayed on fig. 3. as a
 function of the spectator particle ($^{5}$He{} ) momentum.
 In the plane wave limit this would
  correspond to the momentum distribution of 
 $\alpha${} in $^{9}$Be{} if the half-off-shell cross section was
 constant.

\begin{figure}
\resizebox{0.5\textwidth}{!}{
  \includegraphics{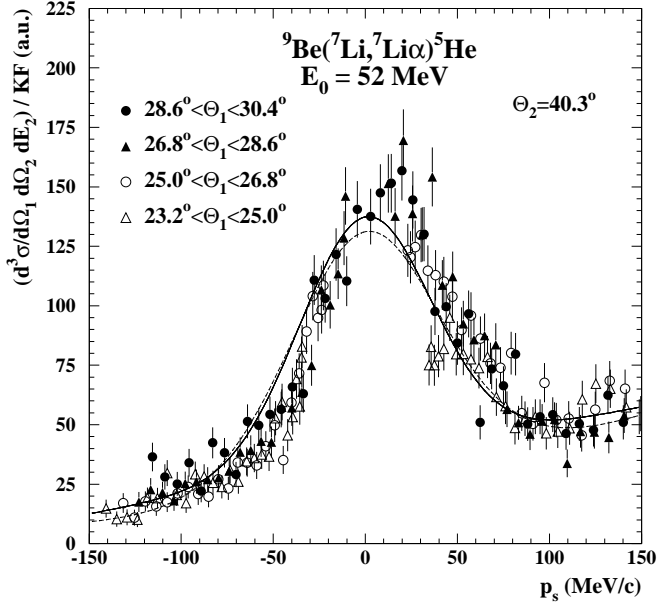}
}
\caption{Plots of
 (d$^{3}\sigma$ / d$\Omega_{^{7}Li}$ d$\Omega_{\alpha}$ dE$_{\alpha}$)/KF
 versus $^{5}$He momentum with $\Theta_{\alpha}$ fixed and four separate 
 $^{7}$Li angles. The curves represent PWIA calculations with
 radial wave functions being a Gaussian (solid line) and Hankel function
 with R$_{c}$ = 4.8 fm (dotted line).  }
\label{fig:3}       
\end{figure}

  The full width at half maximum of the "momentum distribution" is
 around 90 MeV/c to be compared with the values extracted from 
 different quasi-free measurements on $^{9}$Be{}, ranging from 
 70 to 130 MeV/c. The maximum of the distribution is shifted with
 respect to the p$_{s}$ = 0 point. The lower width and the shift 
 could be attributed to strong distortion effects and the variability 
 of the $^{7}$Li{}-$\alpha${} scattering cross section with 
 energy and angle.

 For comparison are shown the fits to the data using two radial 
 intercluster ($^{4}$He{}-$^{5}$He{}) wave functions (Gaussian -
 solid line, Hankel with a cut-off
 - dotted line) and with a linear "background" added.
 The large radial cut-off (4.8 fm), needed to get agreement with the
 data, can be interpreted, as in other 
 cases before, to be a consequence of the very peripheral nature of
 this process.
 
 In conclusion, strong contributions of the $^{7}$Li{}-$\alpha${}
 quasi-free scattering are observed in the
 $^{9}$Be{}($^{7}$Li{}, $^{7}$Li{}$\alpha${})$^{5}$He{} reaction
 at center-of-mass energy lower than 4.5 MeV per nucleon.
 It could be expected
 that these contributions are present even at lower energies
 in analogy with the behaviour of some other quasi-free processes like
 in the $^{6}$Li{}(d, 2d)$^{4}$He{} \cite{8} and 
 $^{6}$Li{}($^{6}$Li{}, 2$\alpha${})$^{4}$He{} \cite{9} reactions.
 Similarly, one can expect that other quasi-free scattering
 contributions, like $^{7}$Li{}-$^{5}$He{}, $^{9}$Be{}-$\alpha${}
 and $^{9}$Be{}-t, play an important role
 in the $^{7}$Li{}+$^{9}$Be{} collisions, too. These processes together
 with sequential and other break-up processes then may make a significant
 part of the total reaction cross section. This is in accordance with
 the plausible interpretation \cite{1}, that the inhibition of the 
 complete fusion in collisions between light weakly bound
 nuclei at lower energies has its origin in their "fragility", i.e. high
 probability of their  break-up.\\
 \\
   The authors wish to thank Mr. C. Marchetta for target preparation
 and the staff of the SMP Tandem Van de Graaff accelerator for their
 efforts during the experiment.


\end{document}